\documentclass[%
reprint, 
superscriptaddress, 
showpacs,
amsmath, amssymb, 
aps, 
prb,
]{revtex4-1}
\usepackage{graphicx}
\usepackage{float}
\usepackage{dcolumn}
\usepackage{bm}
\usepackage{hyperref}
\hypersetup{colorlinks=true, citecolor=blue, urlcolor=blue, linkcolor=blue}
\begin{document}
	\title{Valley and Valley-like Split-ring Topological Photonic Crystal}
	\author{Hui-Chang Li} \author{Chen Luo} \author{Tai-Lin Zhang} \author{Jian-Wei Xu} \author{Xiang Zhou} \author{Yun Shen} \email{shenyun@ncu.edu.cn}
	\affiliation{Department of Physics, Nanchang University, Nanchang 330031, China}  
	\author{Xiao-Hua Deng} \email{dengxiaohua0@gmail.com}
	\affiliation{Institute of Space Science and Technology, Nanchang University, Nanchang 330031, China}
	

	\date{\today}
	\begin{abstract}
		
		In the research of topological phases of matter, valley pseudospins have been introduced into photonic systems. Here, we construct a split-ring photonic crystal (SPC) in which the spilt rings are distributed according to the Kagome model. By rotating three split rings as a whole under the condition of ensuring the existence of $C_{3v}$ symmetry, we obtain a traditional two-band-inversion valley topology (2IVT) driven by opening twofold Dirac degeneracy point. When three split rings are rotated as a whole without ensuring the existence of $C_{3v}$ symmetry, a valley-like topology driven by opening twofold degeneracy point will exist. In particular, when three split rings are rotated separately, three-band-inversion valley-like topology (3IVT) will exist which is also driven by opening twofold degeneracy point. Valley topology and valley-like topology can be described by non-trivial Wannier band (WB) and bulk polarization (BP), and they both have the positive and negative refraction along the Zigzag domain-wall. Our research can be extended to other models, using controllable geometry to construct a variety of topological structures, so as to provide ideas for the research of new topological states.
		
	\end{abstract}
	\maketitle
	
	\section{\label{1}Introduction}
	
	Recent years, the topology in condensed matter physics is combined with classical systems. New topological structures have already sprung up and numerous multifunctional topological structures beneficial to information communication and energy transmission have been realized~\cite{1-1,1-2,1-3,1-4,1-5,1-6}. The concept of valley comes from valley electronics ~\cite{2-1,2-2,2-3,2-4}, which refers to the quantum states of the energy extreme in the momentum space. The quantum valley Hall effect is produced by the opposite magnetic momentum produced by the angular momentum of the wavefunction at the inequivalent valley. Valley chiral edge states in opposite directions can be propagated along the domain-wall without intervalley scattering~\cite{3-1}.
	
	By referring to the classical model~\cite{4-1,4-2,4-3} in condensed matter (such as SSH, Kagome, Honeycomb, etc.) and introducing the symmetry of geometric structure in classical system, numerous valley structures supporting robust transmission and continuous backscattering have been realized in fields such as photonics~\cite{5-1,5-2,5-3,5-4,5-5,5-6,5-7,5-8,5-9,5-10,5-11,5-12}, acoustics~\cite{7-1,7-2,7-3,7-4,7-5,7-6,7-7,7-8,7-9,7-10,7-11,7-12,7-13} and mechanics~\cite{6-1,6-2,6-3,6-4,6-5,6-6}.

	In photonic valley topology transmission, topological refraction in the domain-wall composed of two kinds of valley topology photonic crystals (VTPCs) has also been extensively studied~\cite{8-1}. Along the domain-wall, the direction of edge states projected into the $K$ and $K'$ valley is locked, which can be used to  design unidirectional-propagation waveguides. It's also interesting to note that the edge states projected by $K$ and $K'$ valley exhibit different types of refraction, and directions of the refraction can be determined by the phase-matching condition at the termination of VTPCs.

	In this paper, we construct a SPC in which the spilt rings are distributed according to the Kagome model. In ~\ref{2}, by rotating three split rings as a whole under the condition of ensuring the existence of $C_{3v}$ symmetry, we obtain a traditional 2IVT driven by opening twofold Dirac degeneracy point. The topology of the bandgaps can be described by two different non-trivial WB and BP. Then we use the phase-matching condition to demonstrate the positive and negative refraction phenomenon along the Zigzag domain-wall constructed by 2IVT. In ~\ref{3}, we propose that when three split rings are rotated as a whole without ensuring the existence of $C_{3v}$ symmetry, a valley-like topology which opens twofold degeneracy point will exist. In particular, when three split rings are rotated separately, 3IVT will exist. 3IVT can still be described by nontrivial WB and BP, and has the positive and negative refraction properties of traditional Valley topology. Finally, relevant discussions and conclusions are drawn in ~\ref{4}.

	

	\section{VALLEY SPLIT-RING PHOTONIC CRYSTAL }\label{2}
	In this part, we introduce the proposed structure of valley SPC and the theory of BP and WB needed to characterize the topology. Furthermore, the phenomenon of Valley topological refraction is introduced.


	\subsection{Model}\label{2.1}
	
\begin{figure}
		\centering
		\includegraphics[width=8.6cm]{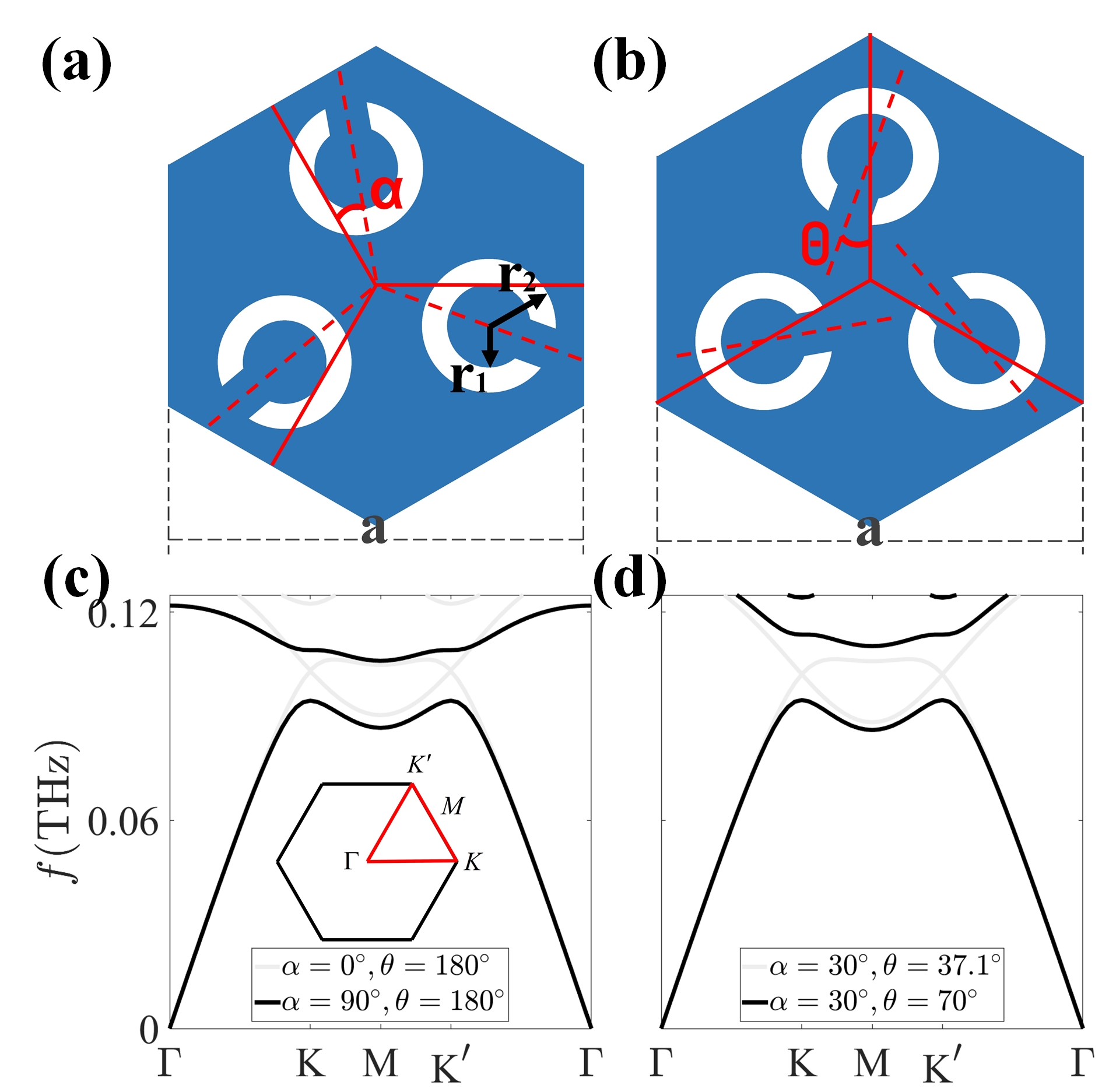}
		\caption{Schematic of the SPC and the energy band diagram. In the case of rotating three spilt rings (a)as a whole and (b) separately, the corresponding rotation angles are $\alpha$ and $\theta$, with the crystal indicated by blue color. The geometric parameters are taken as $a = 400um$, $r_{1}  = 0.1*a$, $r_{2} = 0.16*a$. (c) Energy band diagram of the SPC with $\alpha = 0^{\circ}$, $\theta = 180^{\circ}$(gray solid line) and $\alpha = 90^{\circ}$, $\theta = 180^{\circ}$(black solid line). (d) Energy band diagram of the SPC with $\alpha = 30^{\circ}$, $\theta = 37.1^{\circ}$(gray solid line) and $\alpha = 30^{\circ}$, $\theta=70^{\circ}$(black solid line). The illustration of (c) is the first Brillouin zone(BZ), in which the red triangle marks the path for solving the eigenfrequency.}
		\label{fig:one}
	\end{figure}

	Based on the arrangement of the Kagome model, we designed a SPC. Its unit cell diagram is shown in Fig.~\ref{fig:one}(a)(b). Between split rings and background material is the ideal dielectric conductor boundary, and the relative dielectric constant of the background material is $11.7$. The mode used in this paper is $TM$ mode. The overall rotation angle $\alpha$ and individual rotation angle $\theta$ of the split ring are named, respectively. For 2IVT, as an example, we calculate the energy spectrums of $\alpha = 0^{\circ}$, $\theta = 180^{\circ}$ and $\alpha = 90^{\circ}$, $\theta = 180^{\circ}$(Fig.~\ref{fig:one}(c)). SPC satisfies $C_{3v}$ symmetry in the former case and $C_{3}$ symmetry in the latter case. The reduction of $C_{3v}$ symmetry is the reason for the bandgap of valley topology(~\ref{2.3}). For 3IVT, as an example, we calculate the energy spectrums of $\alpha = 30^{\circ}$, $\theta = 37.1^{\circ}$ and $\alpha = 30^{\circ}$, $\theta = 70^{\circ}$(Fig.~\ref{fig:one}(d)). SPC only satisfies $C_{3}$ symmetry in both cases. At this time, the reason for the bandgap is no longer the reduction of symmetry, and this bandgap have the inversion of states similar to valley topology(~\ref{3.2}).
	
	\subsection{WB AND BP}\label{2.2}
	
	Here, the relevant theory of WB and BP is demonstrated. 
	
\begin{figure}
		\centering
		\includegraphics[width=8.6cm]{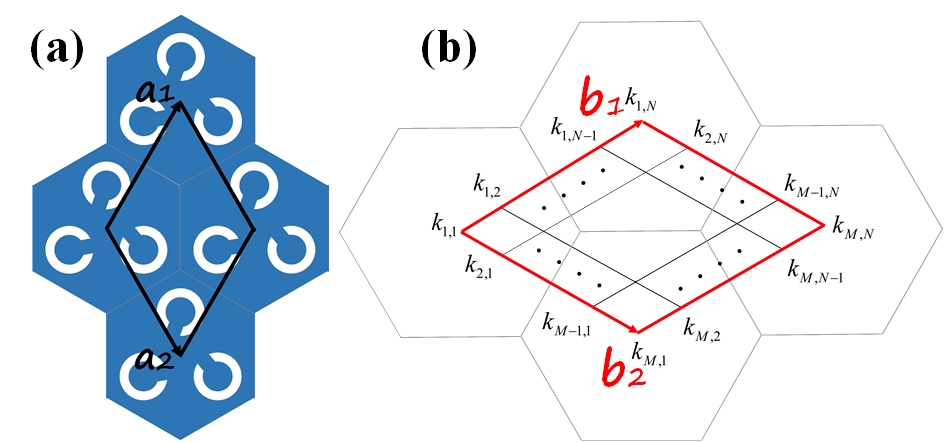}
		\caption{Unit cell and BZ. (a) Unit cell is marked with a black quadrilateral, $\vec{a_{1}}$ and $\vec{a_{2}}$ denote the lattice vectors. (b) BZ is marked with a solid red line, $\vec{b_{1}}$ and $\vec{b_{2}}$ denote the reciprocal lattice vectors. BZ is divided into $M \times N$ regions.}
		\label{fig:two}
	\end{figure}

	First, we define unit cell as shown in Fig.~\ref{fig:two}(a), where $a_{1}=(\frac{1}{2}a,\frac{\sqrt3}{2}a,0)$ and $a_{2}=(\frac{1}{2}a,-\frac{\sqrt3}{2}a,0)$. 
	To consider a unit vector $a_{3} = (0,0,1)$ in the $z$ direction and solve for the reciprocal lattice vector: $b_{1} = \frac{2\pi(a_{2}\times a_{3})}{a_{1}\cdot (a_{2}\times a_{3})}=(\frac{2\pi}{a},\frac{2\pi}{a\sqrt3})$, $b_{2} = \frac{2\pi(a_{3}\times a_{1})}{a_{1}\cdot (a_{2}\times a_{3})}=(\frac{2\pi}{a},-\frac{2\pi}{a\sqrt3})$.

	The corresponding BZ can be obtained as shown in Fig.~\ref{fig:two}(b). The BZ is divided into $M \times N$ $k$ points along $b_{1}$ and  $b_{2}$ directions respectively(we use $M = N = 20$ in all calculations). Then, the Wilson-loop can be defined:
	
	\begin{equation}
		W_{\omega,\upsilon}=F_{\omega,\upsilon}(k) \cdot F_{\omega,\upsilon}(k+\triangle k) \cdot \cdot \cdot F_{\omega,\upsilon}(k+(N-1)\triangle k)
		\label{eq:one}
	\end{equation}

	where 
	\begin{equation}
		F_{\omega,\upsilon}(k)=
		\left(\begin{array}{cccc}
			\Re_{1,1}(k)&\Re_{1,2}(k)&\cdots&\Re_{1,\upsilon}(k) \\
			\Re_{2,1}(k)&\Re_{2,2}(k)&\cdots&\Re_{2,\upsilon}(k) \\
			\vdots&\vdots&\ddots&\vdots \\
			\Re_{\omega,1}(k)&\Re_{\omega,2}(k)&\cdots&\Re_{\omega,\upsilon}(k))
			\label{eq:two}
		\end{array}	\right)
	\end{equation}
	
	$\Re_{\omega,\upsilon}(k) = \left\langle\varphi_{\omega}(k)|\partial_{k}|\varphi_{\upsilon}(k+\triangle k)\right\rangle$ is the Wilson-loop element, in which $\triangle k=\frac{2\pi}{N}$, and the $|\varphi_{\omega,\upsilon}(k)\rangle$ is the periodic part of the wave functions of the $\omega$th, $\upsilon$th order band with wave vector $k$.
	
	We then diagonalize the Wilson-loop operator $W_{\omega,\upsilon}|\chi\rangle = exp(i\cdot 2\pi N_{2}(b_{1}))|\chi\rangle$, where $|\chi\rangle$ is the eigenvector which depends on the Wilson-loop, and the phase $N_{2}(b_{1})$ is the element that forms the Wannier band. As an example, we calculate $N_{2}$($N_{2} = N_{1}$ restricted by the $C_{3}$ symmetry) as a function of $b_{1}$ in all calculations.
	
	Now, the bulk polarization can be defined as~\cite{9-1}
	
	\begin{equation}
		BP = \frac{1}{N} \sum_{b_{1}}N_{2}(b_{1})
		\label{eq:three}
	\end{equation}


	\subsection{2IVT}\label{2.3}

\begin{figure}[htb]
		\centering
		\includegraphics[width=8.6cm]{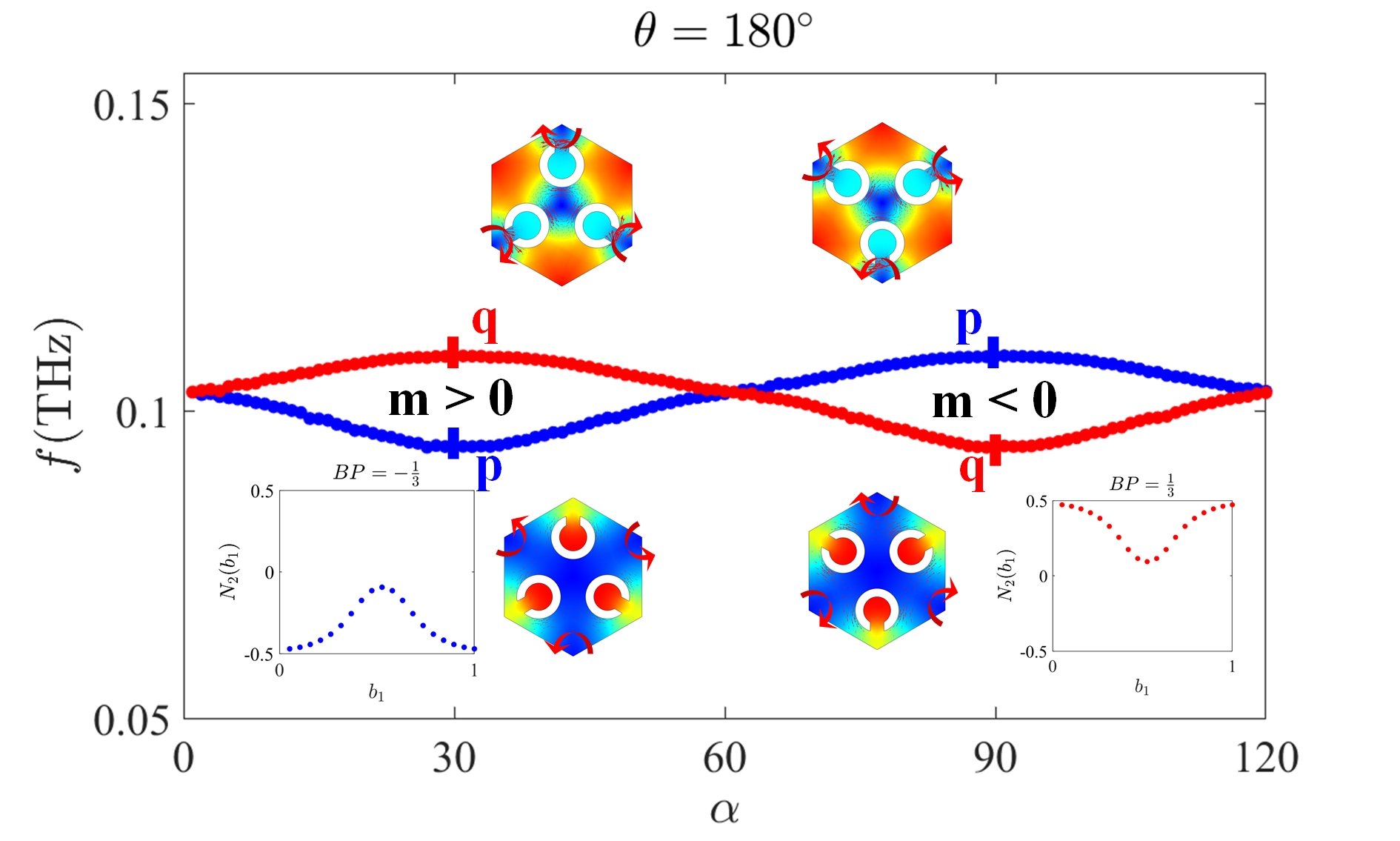}
		\caption{Photonic energy band diagram of 2IVT SPC. The lowest two energy bands at high symmetry point $K$ are calculated as a function of $\alpha$, which is taken here $\theta = 180^{\circ}$. The $WB$, $BP$ and mode distribution of $|TM|$ at $\alpha = 30^{\circ}$ and $\alpha = -30^{\circ}$ are calculated here, which are shown in the illustration (the Poynting vectors represented by the red solid arrows). Blue and red color of dots mark $BP = -\frac{1}{3}$ and $BP = \frac{1}{3}$, respectively.}
		\label{fig:three}
	\end{figure}

	Next, we consider the case of overall rotation of three split rings, and analyze the valley topology of 2IVT.
	First, we analyze the case when $\theta = 180^{\circ}$. By rotating the split rings, characterized by the rotation angle $\alpha$, the original $C_{3v}$ symmetry at $\alpha = 0^{\circ}$ is reduced to the $C_{3}$ symmetry. Therefore, twofold Dirac degeneracy at the symmetry point $K$ and $K'$ is lifted and a bandgap is opened. Here, we refer to the SPC with $\alpha = 30^{\circ}$ as up-triangular SPC (USPC) and $\alpha = 90^{\circ}$ as down-triangular SPC (DSPC). By analyzing the lowest band gap, DSPC and USPC bring distinct topological phases. It is observed that DSPC and USPC have opposite vortices in $p$-state and $q$-state at high symmetry points $K$ and $K'$(Illustration of Fig.~\ref{fig:three}(a)). When the degeneracy is lifted, for DSPC and USPC, the corresponding frequencies of $p$-state and $q$-state are fliped (quantified by Dirac mass m), indicating that the typical band inversion is related to the topological phase transition. Note that the period of the phase diagram is $120^{\circ}$, which is due to the $C_{3}$ symmetry of SPC.


	\subsection{TOPOLOGICAL REFRACTION}\label{2.4}
	
	Next, we demonstrate the topological refraction of the valley edge states from the Zigzag domain-wall into the ordinary crystal space at the termination. 
	
\begin{figure*}[htb]
		\centering
		\includegraphics[width=15cm]{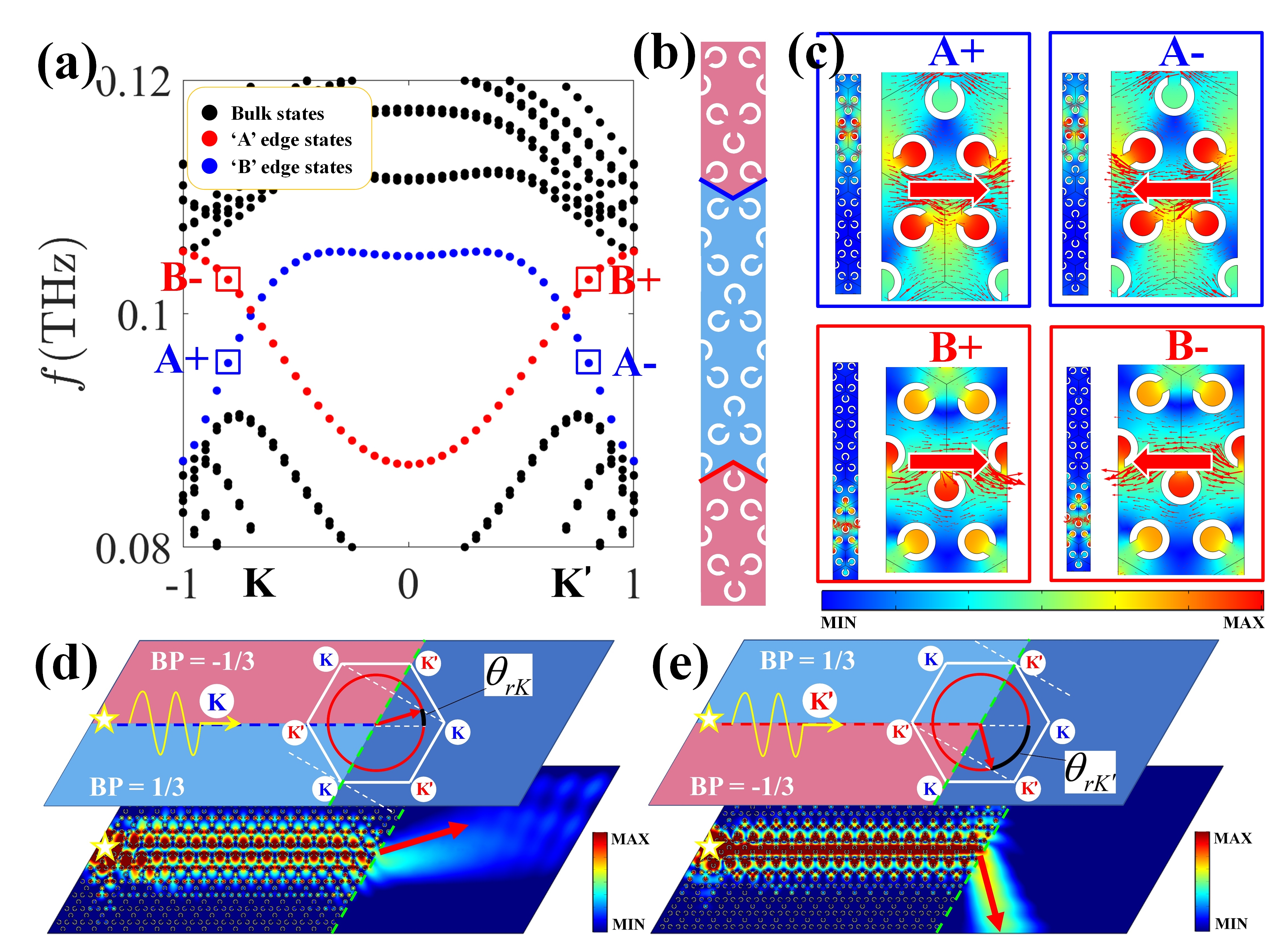}
		\caption{(a) The energy band diagram of supercell(b) composed of USPC and DSPC is simulated. Red and blue dots mark the valley edge states at two different domain-walls ($A$ states and $B$ states), and black dots mark the bulk states. (c) is the mode distribution of $|TM|$ at $A_{+}$, $A_{-}$, $B_{+}$, $B_{-}$ in (a), and the red solid arrows mark the direction of the Poynting vectors. $k$-space analysis on the outcoupling of (d)$K$/(e)$K'$ projected valley edge states along the $A/B$ domain-wall (represented by blue/red solid lines respectively). White solid hexagon represents the first BZ and the red solid circle shows the dispersion in silicon. The termination is marked by a dashed green line and the source is marked by tentagram. The yellow and red solid arrows mark the propagation of the light beam in SPC Zigzag domain-wall and silicon, respectively. Bottom panel: the simulated propagation of light beam through the Zigzag domain-wall with f = $0.102THz$(the red arrow indicates the propagation direction of the light beam in silicon).}
		\label{fig:four}
	\end{figure*}

	By constructing the supercell as shown in Fig.~\ref{fig:four}(b), setting the upper-lower and left-right boundaries as periodic boundaries, and calculating the corresponding supercell band structure (Fig.~\ref{fig:four}(a)). We find that there are two valley edge states at two different domain-walls, which are called $A$ states (at USPC/DSPC) and $B$ states (at DSPC/USPC). For the two valley edge states, their positive/negative slope corresponds to the direction of the Poynting vectors is right/left.

	Then we calculate the propagation characteristics of light beam in a SPC Zigzag domain-wall system (as shown in Fig.~\ref{fig:four}(d)(e)). The direction of the light beam passing through the Zigzag domain-wall depends on the valley type ($K$ or $K'$) from which the edge states are projected~\cite{8-1}. 
	
	We set the light source with frequency of $0.102THz$ on the left side of the domain-wall system composed of DSPC and USPC. 
	
	For the case where the edge states are projected into the $K$ Valley, as shown in the bottom panel of Fig.~\ref{fig:four}(d)(e), SPCs with $BP = \frac{1}{3}$ and $BP = -\frac{1}{3}$ are respectively above and below the domain-wall, and the propagation of the outgoing light beam shows the propagation characteristics of positive refraction. In addition, for the case where the edge states are projected into the $K'$ valley, the propagation of the outgoing light beam shows the propagation characteristics of negative refraction.

	To interpret this phenomenon, the phase-matching condition at the terminal is utilized to calculate the refractive angle here.

	Firstly, we introduce the first BZ and the equifrequency contour of ordinary materials(silicon is used here) [white solid regular hexagons and red solid circles in the illustration of Fig.~\ref{fig:four}(d)(e)] to represents the relative values of the incident vectors(denoted as $K$) and refracted wave vectors (denoted as $k$). 
	
	For the edge states that projected from the $K$ valley, the magnitude of wavevector is $|K|= \frac{2}{3} \cdot \frac{2\pi}{a}$ with the lattice constant $a$ in the reciprocal space. On the flip side, the equifrequency curves in silicon plate can be determined by $|k| = 2\pi\cdot \frac{f\cdot n_{Si}}{c }$, where $f$ represents the incident frequency, $c$ is the light speed in air and $n_{Si}$ is the refractive index of silicon (Here we take $f = 0.102 \times 10^{12} s^{-1}$, $c = 3\times10^{8} m/s$ and $n_{Si} = 3.42$). According to the phase-matching condition to the terminal parallel to $e_{term}$ which satisfies $k\cdot e_{term} = K\cdot e_{term}$. The wavevector in the waveguide satisfies $|k|\cdot \cos(60^{\circ}-\theta_{rK}) = |K|\cdot cos{60^{\circ}}$. It can be calculated that $\theta_{rK} = 15.8^{\circ}$, which is the positive refraction angle of outgoing light beam in the bottom panel of Fig.~\ref{fig:four}(d).
	
	For the edge states that projected from the $K'$ valley, the wavevector in the waveguide is satisfies $|k|\cdot \cos(120^{\circ}+\theta_{rK'}) = |K|\cdot cos{60^{\circ}}$. It is calculated that $\theta_{rK'} = -75.8^{\circ}$, which is also consistent with the negative refraction angle of the outgoing light beam in the bottom panel of Fig.~\ref{fig:four}(e).


	\section{VALLEY-LIKE TOPOLOGICAL SPLIT-RING PHOTONIC CRYSTAL}\label{3}
	
	In this section, we will introduce valley-like SPC topology. The valley-like SPC will no longer realize band inversion by the reduction of $C_{3v}$ symmetry, but rotating the split ring as a whole and separately without ensuring the existence of $C_{3v}$ symmetry.


	\subsection{VALLEY-LIKE TOPOLOGY IN 2IVT}\label{3.1}
	
	Here, we analyze the situation in 2IVT when $\theta~=0^{\circ} or 180^{\circ}$. When the split rings are rotated as a whole, SPC will no longer have $C_{3v}$ symmetry, but always maintain $C_{3}$ symmetry. In particular, at this time, there still exist the ‘open-close-open’ of the bandgap and the band inversion.
	
\begin{figure*}[htb]
		\centering
		\includegraphics[width=15cm]{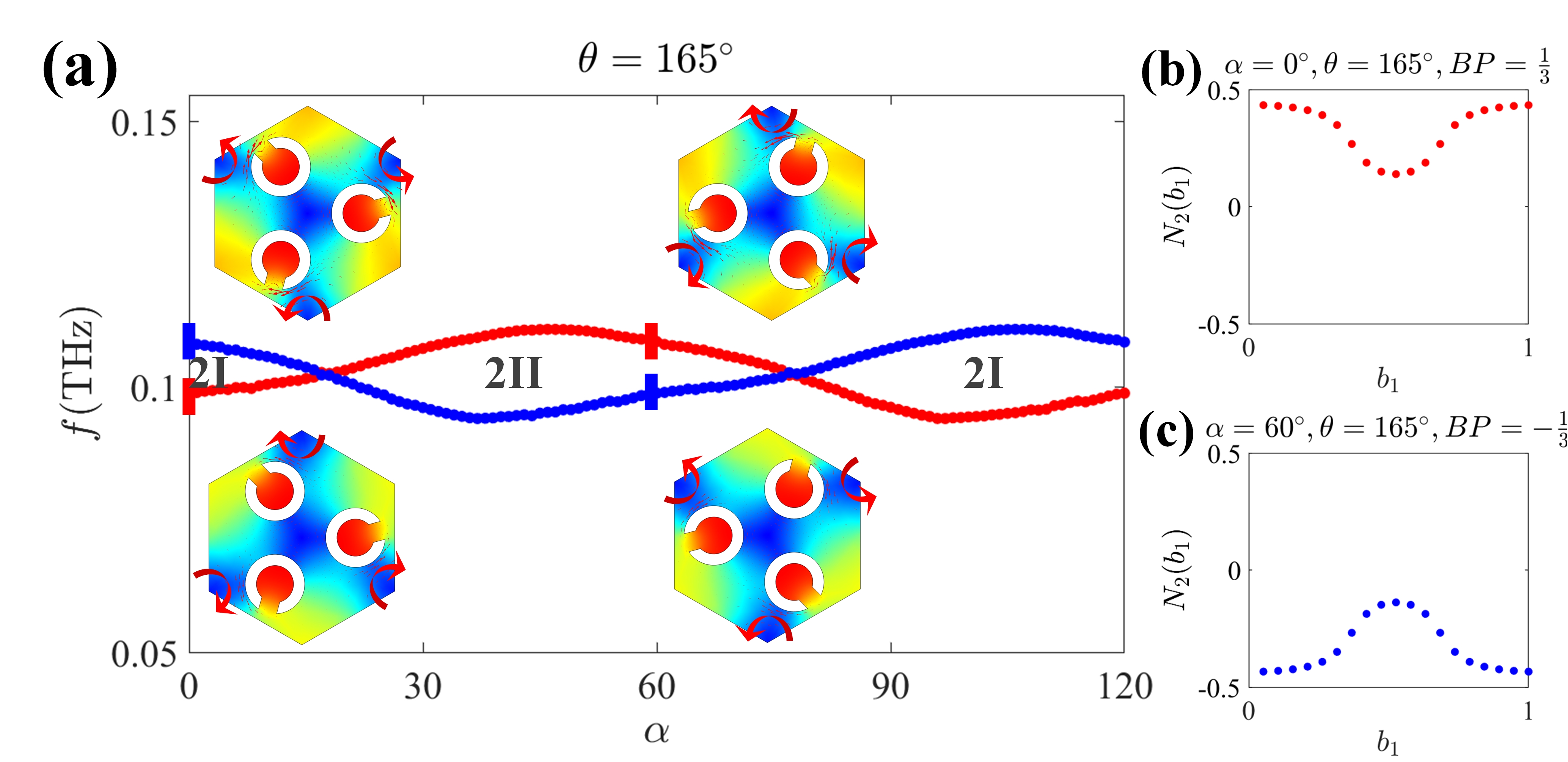}
		\caption{Energy band diagram and WB of SPC. (a) The lowest two energy bands at high symmetry point $K$ are calculated as a function of $\alpha$, which is taken here $\theta = 165^{\circ}$. The mode distribution of $|TM|$ at $\alpha = 0 ^{\circ}$ and $\alpha = 60 ^{\circ}$ are calculated here, which are shown in the illustration (the Poynting vectors represented by the red solid arrows). (b) Wannier band at $\alpha = 0 ^{\circ}$. (c) Wannier band at $\alpha = 60 ^{\circ}$.  Blue and red color of dots mark BP = -$\frac{1}{3}$ and BP = $\frac{1}{3}$, respectively.}
		\label{fig:five}
	\end{figure*}

	As shown in Fig.~\ref{fig:five}(a), when $\theta = 165^{\circ}$, we calculate the corresponding energy spectrum. With the change of $\alpha$, it is observed that the bandgap between the lowest two energy bands exists ‘open-close-open’ similar to the traditional valley topology in~\ref{2.3}, and the corresponding states also reverse. Further, we calculated the BP of the two bandgaps ($\alpha = 0^{\circ}$ and $\alpha = 60^{\circ}$), the correspondence between the value of BP and the position of vortices (at the high symmetry points $K$ or $K'$)is consistent with the valley topology in~\ref{2.3}, which further proves the similarity with valley topology.

	\subsection{VALLEY-LIKE TOPOLOGY IN 3IVT}\label{3.2}
	
	Next, we consider the case of rotating three split rings separately.
	
\begin{figure*}[htb]
		\centering
		\includegraphics[width=15cm]{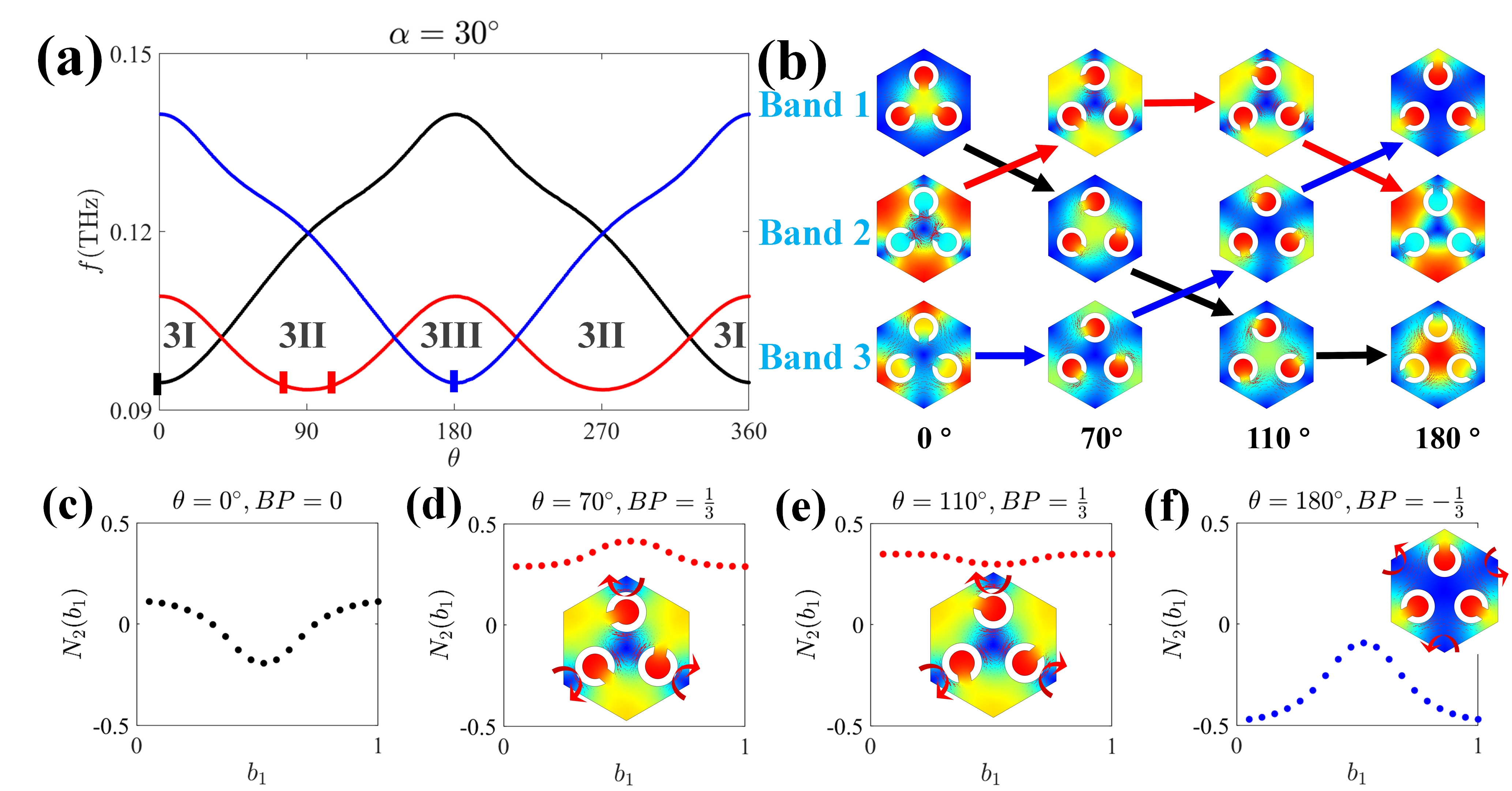}
		\caption{Energy band diagram, mode distribution of $|TM|$ and WB of SPC. (a) The lowest three energy bands at high symmetry point $K$ are calculated as a function of $\theta$, which is taken here $\alpha = 30^{\circ}$. (b) The mode distribution of $|TM|$ in the case of $\theta = 0 ^{\circ}$, $\theta = 70 ^{\circ}$, $\theta = 110 ^{\circ}$ and $\theta = 180 ^{\circ}$ are calculated which shows the inversion of states between the lowest three energy bands. Wannier band (c) at $\theta = 0 ^{\circ}$, (d) at $\theta = 70 ^{\circ}$, (e) at $\theta = 110 ^{\circ}$, (f) at $\theta = 180 ^{\circ}$, black, blue and red color of dots mark BP = 0, BP = -$\frac{1}{3}$ and BP = $\frac{1}{3}$, respectively. The illustration in (d)-(f) is the corresponding mode distribution of $|TM|$, and the Poynting vectors represented by the red solid arrows.}
		\label{fig:six}
	\end{figure*}

	Firstly, we analyze the case when $\alpha = 30^{\circ}$. Under this circumstances, by rotating SPC characterized by angle $\theta$, we calculate the energy spectrum composed of the lowest three bands at point $K$. It can be seen that there doesn't exist twofold Dirac degeneracy point at the angle ($\theta = 0^{\circ}$ and $\theta = 180^{\circ}$) when SPC satisfies $C_{3v}$ symmetry, but rotating the split rings will still lift the degeneracy of twofold point at other angle. In particular, we find that in this case, the inversion of the states occurs between the lowest three energy bands (as shown in Fig.~\ref{fig:six}(b)). During the inversion of the states, the bandgaps between the first and second energy bands are divided into three types: 3I, 3II and 3III, and the values of BP are: $0$, $\frac{1}{3}$ and -$\frac{1}{3}$ respectively. Through the division of BP, 3II and 3III correspond to the valley topological bandgaps of $m<0$ and $m>0$ in Fig.~\ref{fig:three}(a) respectively.

	The mode distributions of $|TM|$ of the first energy band in the case of (3I)$\theta = 0^{\circ}$, (3II)$\theta = 70^{\circ}$, (3II)$\theta = 110^{\circ}$, and (3III)$\theta = 180^{\circ}$ are shown in Fig.~\ref{fig:six}(c)-(f), and the correspondence between the value of BP and the position of vortices (at the high symmetry points $K$ or $K'$) is consistent with the valley topology in~\ref{2.3}, which further proves the similarity with valley topology. 
	
	\subsection{VALLEY AND VALLEY-LIKE EDGE STATES AND TOPOLOGICAL REFRACTION} \label{3.3}
	\begin{figure*}[htb]
		\centering
		\includegraphics[width=15cm]{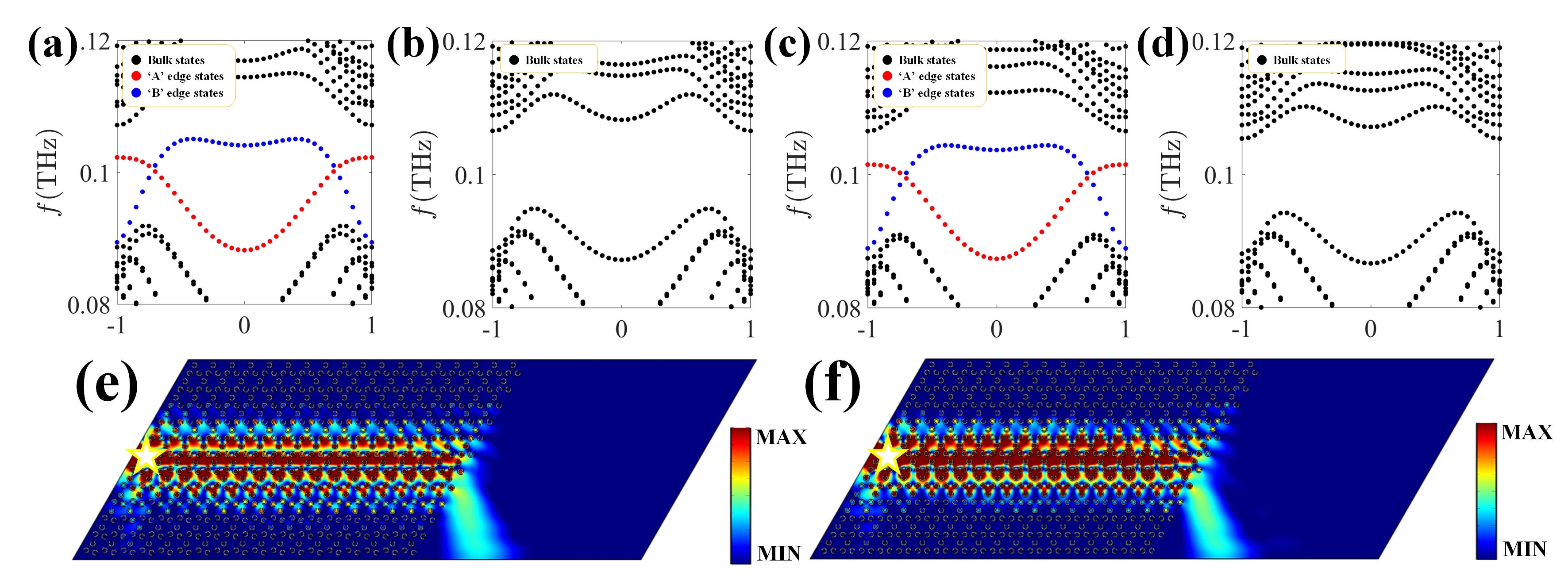}
		\caption{The energy band diagram of supercell with (a) $\alpha = 30^{\circ}+\theta = 90^{\circ} || \alpha = 45^{\circ}+\theta = 165^{\circ}$, (b) $\alpha = 30^{\circ}+\theta = 90^{\circ} || \alpha = 105^{\circ}+\theta = 165^{\circ}$, (c) $\alpha = 30^{\circ}+\theta = 90^{\circ} || \alpha = 30^{\circ}+\theta = 180^{\circ}$, (d) $\alpha = 30^{\circ}+\theta = 90^{\circ} || \alpha = 90^{\circ}+\theta = 180^{\circ}$ is calculated. Red and blue dots mark the edge states at two different domain-walls. (e) The simulated propagation of light beam through the zigzag domain-wall in the case of (a) with $f = 0.101THz$. (f) The simulated propagation of light beam through the Zigzag domain-wall in the case of (c) with $f = 0.101THz$.}
		\label{fig:seven}
	\end{figure*}
	It should be noted that for the valley-like topological photonic crystal, the properties of refraction along the Zigzag domain-wall are consistent with the traditional valley topology (~\ref{2.4}). The key point is that domain wall needs to be composed of crystals with two different non-trivial BP.
	
	Some special cases are selected to verify the properties of valley-like 3IVT and 2IVT similar to valley topology transmission anomalous refraction (Fig.~\ref{fig:seven}). 
	
	Here, we choose $\alpha = 30^{\circ}+\theta = 90^{\circ}$ case with BP = $\frac{1}{3}$ to form the domain-wall together with the other four cases, and calculate their supercell energy band diagrams. The four cases selected are: 1: $\alpha = 45^{\circ}+\theta = 165^{\circ}$, 2: $\alpha = 105^{\circ}+\theta = 165^{\circ} $, 3: $\alpha = 30^{\circ}+\theta = 180^{\circ}$, 4: $\alpha = 90^{\circ}+\theta = 180^{\circ}$, respectively. It is verified that different BP in the domain-wall system is the key to the existence of edge states and topological refraction.


	\section{CONCLUSION}\label{4}
	
	In summary, we propose a topological photonic crystal driven by rotating the split rings, which has the traditional valley topology (two-band-inversion driven by lifting twofold Dirac degeneracy point) and valley-like topology (two-band-inversion and three-band-inversion by lifting twofold degeneracy point). Their topological properties can be described by non-trivial WB and BP. Along the Zigzag domain-wall which composed of VTPCs with different BP, there exist positive and negative refraction, which can be described by the phase-matching condition between VTPCs and ordinary crystal, and the corresponding refraction angle can be calculated. Our research has made a breakthrough in valley-like topology crystals which do not strictly meet the traditional valley topology. Furthermore, the multi-band-inversion existing in valley-like topology is expected to be realized by other models.
	
	\begin{acknowledgments}
		Acknowledgments: This work was supported by the National Natural Science Foundation of China (Grant numbers 61865009, 61927813).
	\end{acknowledgments}




\end{document}